\documentclass{osa-article}

\journal{oe}

\usepackage{CJK}
\usepackage{amsmath}
\usepackage{bm}
\usepackage[Symbol]{upgreek}
\articletype{Research Article}

\begin{document}

\title{Tuning toroidal dipole resonances in dielectric metamolecules by an additional electric dipolar response}

\author{Tiancheng Huang\authormark{1}, Boxiang Wang\authormark{1}, and Changying Zhao\authormark{1,*}}

\address{\authormark{1}Institute of Engineering Thermophysics, Shanghai Jiao Tong University, Shanghai 200240, China}

\email{\authormark{*}changying.zhao@sjtu.edu.cn} 



\begin{abstract}
With the rise of artificial magnetism and metamaterials, the toroidal family recently attracts more attention for its unique properties. Here we propose an all-dielectric pentamer metamolecule consisting of nano-cylinders with two toroidal dipolar resonances, whose frequencies, EM distributions and Q factor can be efficiently tuned due to the additional electric dipole mode offered by a central cylinder. To further reveal the underlying coupling effects and formation mechanism of toroidal responses, the multiple scattering theory is adopted. It is found that the first toroidal dipole mode, which can be tuned from 2.21 to 3.55 $\upmu$m, is mainly induced by a collective electric dipolar resonance, while the second one, which can be tuned from 1.53 to 1.84 $\upmu$m, relies on the cross coupling of both electric and magnetic dipolar responses. The proposed low-loss metamolecule and modes coupling analyses may pave the way for active design of toroidal responses in advanced optical devices.            
\end{abstract}

\section{Introduction}\label{sec1}
 The toroidal multipoles, as the counterpart of electric and magnetic multipole families, have been always neglected in previous studies for their weak or ambiguous resonances in natural materials. However, due to the rapid development in both nano-fabrication technique \cite{KSDG17, PaZL00} and artificial magnetism realized by metamaterials \cite{JaJa16, HuWZ18}, toroidal dipole (TD), which is the most fundamental mode in its family, can be observed both in metallic and dielectric subwavelength metamolecules. Toroidal multipoles offer a new route to manipulate light scattering by subwavelength dielectric particles, which is conventionally based on the tuning of electric and magnetic multipole families \cite{ElTM14, lin2015mode, doi:10.1021/nn301398a}. The near field resonance of TD can be regarded as currents flowing on the torus surface along its meridians, which is instinctively different from an electric dipole (ED) or magnetic dipole (MD). However, TD and ED share the same far-field scattering patterns and parity properties, which can realize a destructive interference and create a radiationless source (anapole mode) to produce oscillating and propagating vector potential without electromagnetic fields \cite{PhysRevX.5.011036}. Besides, TD can support many other intriguing phenomenon, such as the ideal magnetic dipole \cite{feng2017ideal}, optical activity \cite{papasimakis2009gyrotropy}, optical transparency \cite{fedotov2013resonant,LiZM15}, and ultra-high energy density \cite{Wei:16,doi:10.1021/acsphotonics.7b01440}. Meanwhile, TD has been widely implemented to enhance light-matter interactions like all-optical Hall effect \cite{dong2013all}, non-linear laser generation \cite{shorokhov2016multifold}, and optical force\cite{ZWLS15}. Thus, TD mode is of great importance in the thriving fields of metamaterials and nanophotonics.  

The tuning of resonance modes in metamaterials is significant in practical applications, where broad tunable frequency ranges are required to cover more situations. Much work so far has been focused on the build of metamolecules with distinct TD modes, when little attention has been paid to the adjustment of resonance frequencies and strength of TD modes. In researches on metallic split-rings \cite{dong2012toroidal} and dumbbell-shaped-aperture \cite{fedotov2013resonant} metamaterials, resonance frequencies and Q factor of TD modes have been investigated with the change of fold number in their rotation symmetrical metamolecules, when both Q factor and resonance frequencies were improved with the increment of fold number. In the same time, overall geometry factors, including gap distance \cite{liu2017high,dong2013all,li2017super}, particle separation \cite{liu2017high,TTKS16}, aspect ratio \cite{ospanova2018anapole,terekhov2017resonant,ZENY17}, and the ratio of outer radius to inner radius \cite{LSWZ15,liu2015efficient,LiZM15,liu2015invisible} (in core-shell structures), is another common way to tune TD or other resonance modes in metamaterials. Nevertheless, most of these investigations are dedicated to describing the phenomenon when one of these geometry factors is changed, which is not an active tuning of TD modes since the underneath mechanism is seldom well-discussed. A theoretical study \cite{WJRF16} on the toroidal dipole moment resulted from excitation eigenmodes of interacting plasmonic nanorods were conducted using dipole approximation considering only the effect of electric dipoles, which can hardly describe the coupling in dielectric materials due to the ignorance of magnetic responses. Another representative comprehensive study \cite{TTKS16} about all-dielectric metamolecules composed by multi-cylinder structures were investigated based on the lumped wire theory and coupled mode theory. Yet the coupled mode theory in this work cannot accurately solve the collective mode frequencies. Consequently, the coupling mechanism of TD resonances in dielectric metamaterials demands further investigation to fill the knowledge gap between the observations and theoretical explanations.  
   
In this work, we design a metamolecule consisting of subwavelength silicon cylinders which can support two toroidal dipole modes with large tunable frequency ranges. The Cartesian multipole decomposition method is devoted to displaying scattering contributions of electromagnetic and toroidal multipoles, which can determine the appearance of TD modes. In this way, we find that the frequency tunability of TD modes in the proposed metamolecule is realized by the electric dipolar response provided by the central cylinder. This additional electric dipolar response can also adjust the Q factor and enhance the maximum of electric density through modes coupling effects. After analyzing the near-field coupling of TD modes by multiple scattering theory, we find that the formation mechanism of these TD modes are different. The toroidal mode with wider tunable range is induced mainly by a collective electric dipolar resonance, which then can couple with trivial magnetic dipolar response and form a head-to-tail aligned magnetic current loop. And the other TD with narrower tunable range results from the coupling of trivial electric and magnetic dipolar responses. We hope that the numerical and theoretical results could be helpful to the active design of toroidal dipolar responses with high performance.


\section{Model and Methods}\label{sec2}

\begin{figure}[htbp]
	\centering
	\includegraphics[keepaspectratio,width=0.5\textwidth,height=0.5\textheight]{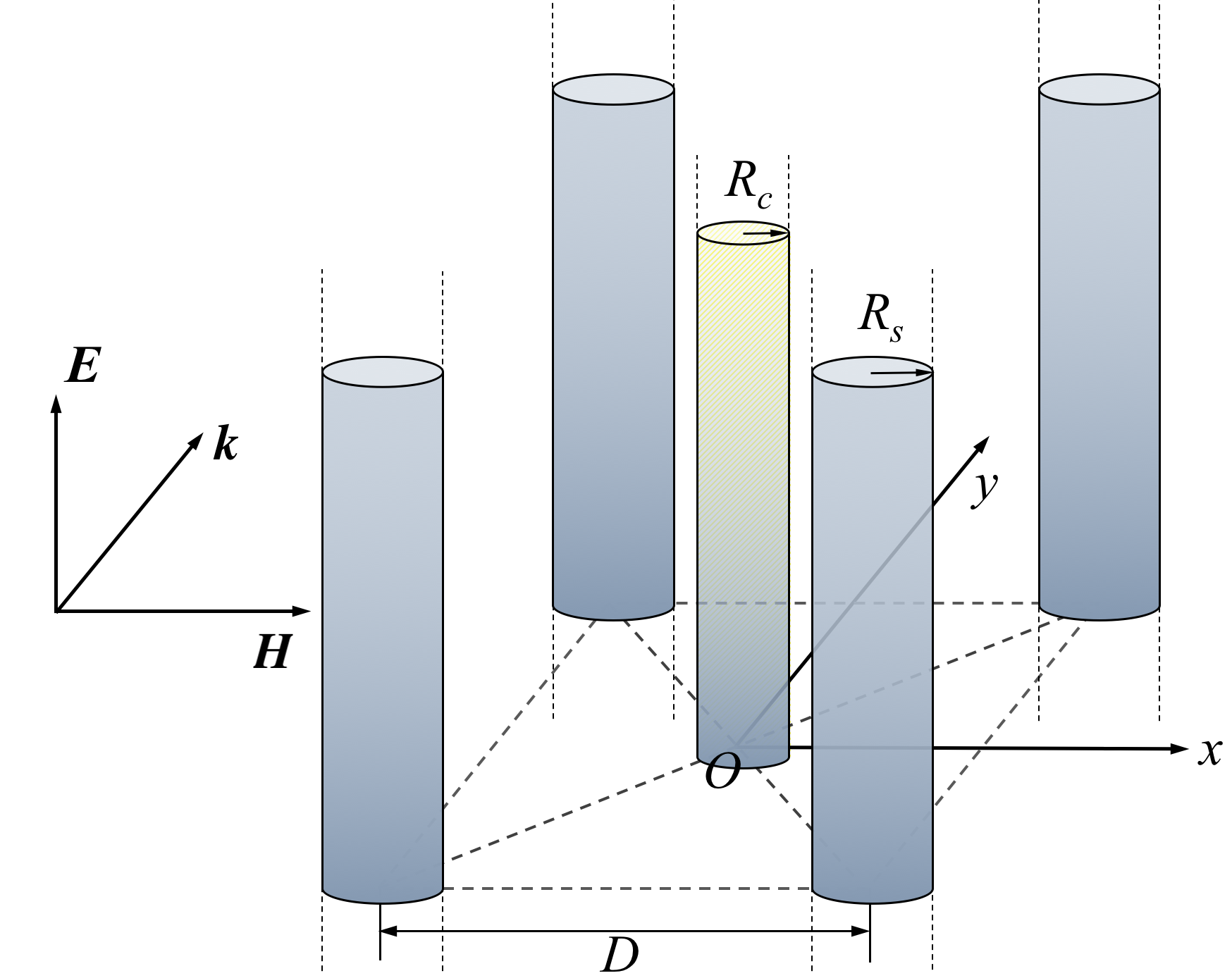}
	\caption{Illustration of the dielectric pentamer consist of infinite-long subwavelength cylinders. $D$ denotes the side length of the lattice. $R_c$ and $R_s$ are radii of the central and surrounding cylinders, respectively.}\label{fig1}
\end{figure}

The geometry structure of pentamer metamolecule composed by infinite-long dielectric subwavelength cylinders is shown in Fig. \ref{fig1}. This pentamer can be divided into two separated parts: one part is the central cylinder and the other part is the rest four identical cylinders, namely the surrounding cylinders. $R_c$ and $R_s$ are radii of the central and surrounding cylinders, respectively. Geometry centers of the surrounding cylinders are located at the corners of a square lattice with side length $D$, when the center of central cylinder overlaps with the center of square lattice. The permittivity of silicon can be regarded as a real constant 12.25 in the near infrared region. An s-polarized plane wave propagating along the y-axis serves as the incident source. Electromagnetic fields and resonant modes of the pentamer are solved by the methods illustrated in this section.

\subsection{Cartesian Multipole decomposition method}\label{sec2_1}
Even though the same far-field radiation patterns can be obtained by both spherical and Cartesian multipole decomposition, the excitation of toroidal multipoles can be directly exhibited under Cartesian multipole expansions. Thus, we here utilize the Cartesian multipole decomposition method to show the scattering contributions of TD mode and other electromagnetic multipoles. To be more accurate, we apply a two-dimensional (2d) Cartesian decomposition conducted in a recent theoretical work \cite{li2018origin}, instead of using a three-dimensional decomposition method in a extended 2d metamolecule\cite{PhysRevX.5.011036}. Besides, the total EM fields are solved by finite-difference time-domain method (FDTD).  

Following Maxwell's equations and Jackson's formulation \cite{jackson2012classical}, the multipole coefficients of s-polarized waves in 2D polar coordinates can be obtained by
\begin{equation}
{ A }_{ m }=\frac { \eta k }{ 4 } \int { {\textbf{J}}_{ z } } { e }^{ im\varphi  }[-{ J }_{ m }]d\textbf{s},\label{eq:1}
\end{equation}
\begin{equation}
{ E }_{ Z }=\sum _{ m=-\infty  }^{ m=+\infty  }{ { A }_{ m }{ H }_{ m }^{ (1) }{ e }^{ im\varphi  } },\label{eq:2}
\end{equation}
where $k$ is the wave number in vacuum, the wave impedance $\eta =\sqrt { { \mu  }_{ 0 }/{ \epsilon  }_{ 0 }{ \epsilon  }_{ r } }$, ${ r }=\sqrt { (x^{ 2 }+{ y }^{ 2 }) } (\cos  ({ \varphi  })\hat { x } +\sin  ({ \varphi  })\hat { y } )$ , the polar angle $\varphi ={ \tan^{ -1 } }(y/x)$, ${ \textbf{J(r)} }$ is the scattering current, defined as ${ \textbf{J(r)} }=-i\omega \epsilon _{ 0 }(\epsilon(\textbf{r})-\epsilon_h )\textbf{E(r)}$ , ${J}_{m}$ is the Bessel function of the first kind with order \emph{m} and argument \emph{kr}, and ${ H }_{ m }^{ (1) }$ is the first kind Hankel function with order \emph{m} and argument \emph{kr}. Thus, we can calculate multipole coefficients under Cartesian coordinates. Due to the Cartesian toroidal dipole has the same radiation pattern as its corresponding electric (magnetic) dipole but scales with two more orders of \emph{kr}, thus the Bessel function is expanded in powers of \emph{kr,}
\begin{equation}
{ J }_{ m }=\frac { 1 }{ m! } (\frac { kr }{ 2 } )^{ 2 }-\frac { 1 }{ (m+1)! } (\frac { kr }{ 2 } )^{m+2}+\cdots.
\label{eq:3}
\end{equation}
The two leading terms are kept to compare with the Cartesian multipole expansion. The first term is denoted as $J_{m,0}$, and the second term is denoted as $J_{m,1}$. In this way, Eq. \ref{eq:3} can be rewritten as
\begin{equation}
A_{m}\approx \frac { \eta k }{ 4 } \int { { J_{ z } }e^{ im\varphi  }[-(J_{ m,0 }+J_{ m,1 })] } d\textbf{s}.
\label{eq:4}
\end{equation}
We know that $J_{m,1}$ is two order higher than $J_{m,0}$, which is the same as the coefficients of electric and toroidal dipoles. Thus, we can separate the toroidal dipole from the conventional electromagnetic multipoles. To be specific, the conventional electric dipole is devided into a Cartesian electric dipole (CED) and a toroidal dipole. Similarly, the conventional magnetic dipole is devided into a Cartesian magnetic dipole (CMD) and a magnetic toroidal dipole (MTD) which is defined to have the same scattering pattern with a magnetic dipole. As shown in Eqs. \ref{eq:5} and \ref{eq:6}, toroidal dipoles can be found in the higher order terms in $A_0$ and $A_1$, respectively. The expansions of Bessel function are not required here since higher-order terms (m>1) contain no toroidal dipole term. 
\begin{equation}
{ A }_{ 0 }=-\frac { \eta k }{ 4 } (1-\frac { (kr)^{ 2 } }{ 4 } )\textbf{J}_{z}d\textbf{s}
\label{eq:5}
\end{equation}
\begin{equation}
{ A }_{ 1 }=-\frac { \eta k }{ 4 } (\frac { kr }{ 2 } -\frac { (kr)^{ 3 } }{ 16 } ){ \textbf{J}_{z} }{ e }^{ i\varphi  }d{ \textbf{s} }
\label{eq:6}
\end{equation}
Then, the scattering cross section can be obtained by summing up the electric and magnetic field of all orders m:
\begin{equation}
C_{ sca,s }=\sum _{ m }{ \frac { 4 }{ k|{ E }_{ 0 }|^{2} }  } |A_{ m }|^{ 2 }.
\label{eq:7}
\end{equation}

\subsection{Multiple scattering theory and coupling effects}\label{sec2_2}

\begin{figure}[htbp]
	\centering
	\includegraphics[keepaspectratio,width=0.5\textwidth,height=0.5\textheight]{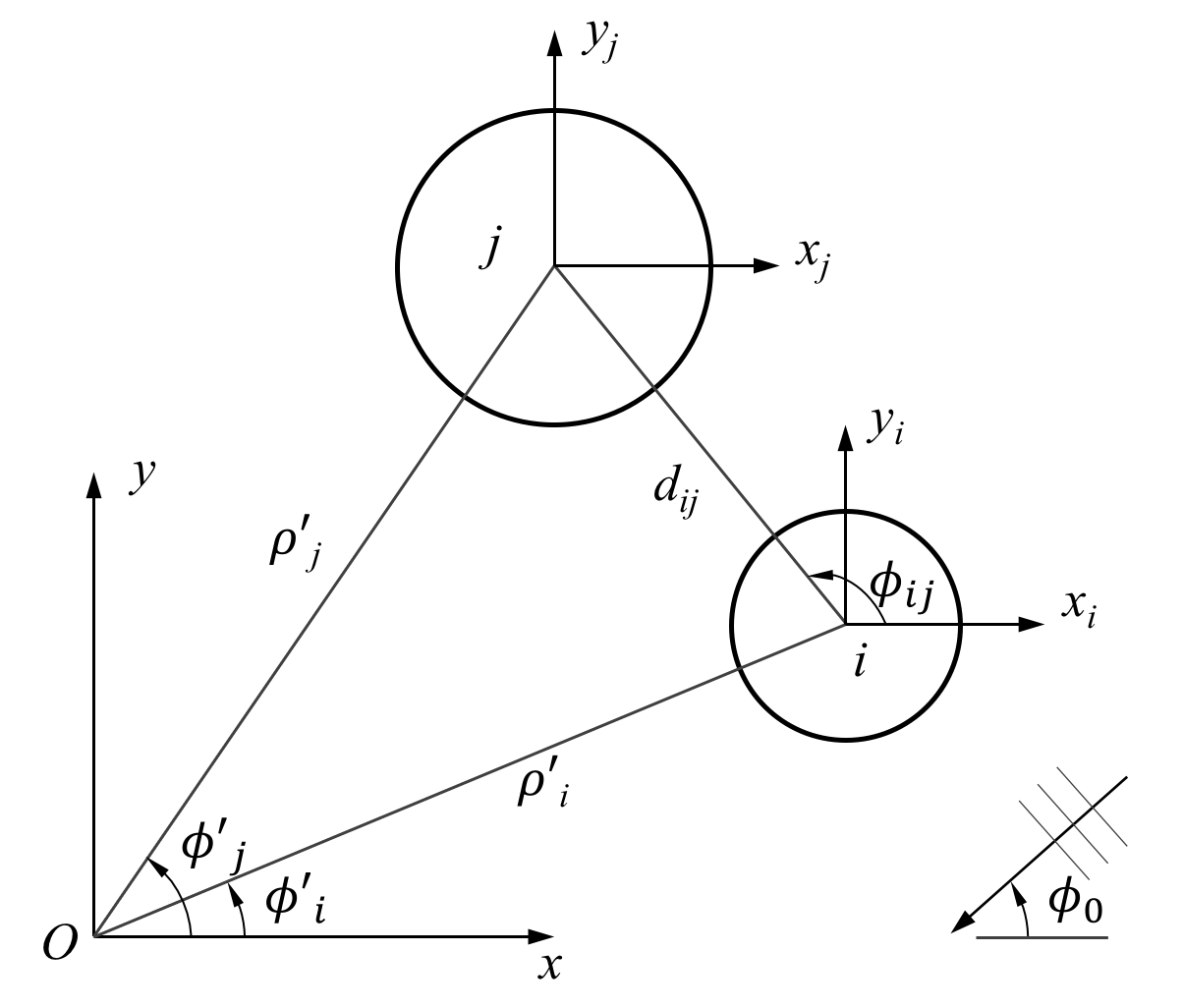}
	\caption{Schematic of multiple cylinders and the coordinate systems.}\label{fig2}
\end{figure}

The coupling of different resonance modes in a 2d multi-cylinders system can be considered in the multiple scattering theory, which makes it a efficient tool to investigate modes coupling effects.  In this method, the modes coupling can be solved based on the vector spherical expansion function \cite{Hulst1957Light}. The geometry structure and parameters of the circular cylinders system are shown in Fig. \ref{fig2}. $i$ and $g$ denote the number of cylinders and $d_{ig}$ is the distance between them. The incident direction of the plane wave is defined by the angle $\phi_0$, and other useful angles are marked in this figure. From the previous studies \cite{123363,doi:10.1002/lpor.201700103,liu2015invisible}, we can obtain the scattering coefficients from the \emph{i}th cylinder by
	\begin{equation}
	{ a }_{ i }^{ l }=\sum _{ g=1 }^{ M }{ \sum _{ n=-\infty  }^{ \infty  }{ { c }_{ gn }[{ b }_{ ig }^{ ln }(1-{ \delta  }_{ ig })+{ d }_{ i }^{ l }{ \delta  }_{ ln }{ \delta  }_{ ig }] }  },
	\label{eq7}
	\end{equation}	
where Kronecker's delta function ${ \delta  }_{ ig }=1$ if $i=g$ and zero otherwise. Parameters $n$ and $l$ are the order of resonance modes. For all cylinders, we can rewrite Eq.(1) in the following matrix form: 		

		\begin{equation}
		\begin{bmatrix} \left[ { A }_{ 1 }^{ l } \right]  \\ \vdots  \\ \left[ { A }_{ i }^{ l } \right]  \\ \vdots  \\ \left[ { A }_{ M }^{ l } \right]  \end{bmatrix}=\begin{bmatrix} \left[ { S }_{ 11 }^{ ln } \right]  & \cdots  & \left[ { S }_{ 1g }^{ ln } \right]  & \cdots  & \left[ { S }_{ 1M }^{ ln } \right]  \\ \vdots  &  & \vdots  &  & \vdots  \\ \left[ { S }_{ i1 }^{ ln } \right]  & \cdots  & \left[ { S }_{ ig }^{ ln } \right]  & \cdots  & \left[ { S }_{ iM }^{ ln } \right]  \\ \vdots  &  & \vdots  &  & \vdots  \\ \left[ { S }_{ M1 }^{ ln } \right]  & \cdots  & \left[ { S }_{ Mg }^{ ln } \right]  & \cdots  & \left[ { S }_{ MM }^{ ln } \right]  \end{bmatrix}\begin{bmatrix} \left[ { C }_{ 1 }^{ l } \right]  \\ \vdots  \\ \left[ { C }_{ i }^{ l } \right]  \\ \vdots  \\ \left[ { C }_{ M }^{ l } \right]  \end{bmatrix},\label{eq8}
		\end{equation}
where the vector $[A]$ and $[C]$ are of dimension $\sum _{ i=1 }^{ M }{ (2{ N }_{ i }+1) }$ and $N_i$ is the order of expansion term which stands for corresponding mode. The sub-matrices $\begin{bmatrix} { S }_{ ig }^{ ln } \end{bmatrix}$ of the coupling matrix $[S]$ can be obtained by

		\begin{equation}
		\left[ { S }_{ ig }^{ ln } \right] =\begin{cases} \left[ { b }_{ ig }^{ ln } \right] \quad \quad ,\quad i\neq g \\  \\ \left[ { d }_{ i }^{ l }{ \delta  }_{ ln } \right] \quad ,\quad i=g \end{cases}
		\label{eq9}
		\end{equation}
where $l, n=0, \pm 1, \pm 2,…, \pm N_i$ , and the mutal interation elements ${ b }_{ ig }^{ ln }$ and self-interaction elements ${ d }_{ i }^{ l }$ are given respectively by	
		\begin{equation}
		{ b }_{ ig }^{ ln }=-{ H }_{ l-n }^{ (2) }({ k }_{ 0 }{ d }_{ ig }){ e }^{ -j(l-n){ \phi  }_{ ig } },
		\label{eq10}
		\end{equation}

		\begin{equation}
		{ d }_{ i }^{ l,s }=\frac{H^{ (2)' }_l({ k }_{ 0 }{ a }_{ i })J_l({ k }_{ i }{ a }_{ i })-{ m }_{ i }H^{ (2) }_l({ k }_{ 0 }{ a }_{ i })J'_l({ k }_{ i }{ a }_{ i })}{{ m }_{ i }J_{ l }({ { k }_{ 0 }{ a }_{ i } })J'_{ l }({ { k }_{ i }{ a }_{ i } })-J'_{ l }({ { k }_{ 0 }{ a }_{ i } })J_{ l }({ { k }_{ i }{ a }_{ i } }) },
		\label{eq11}
		\end{equation}	

		\begin{equation}
		{ d }_{ i }^{ l,p }=\frac{{ m }_{ i }H^{ (2)' }_l({ k }_{ 0 }{ a }_{ i })J_l({ k }_{ i }{ a }_{ i })-H^{ (2) }_l({ k }_{ 0 }{ a }_{ i })J'_l({ k }_{ i }{ a }_{ i })}{J_{ l }({ { k }_{ 0 }{ a }_{ i } })J'_{ l }({ { k }_{ i }{ a }_{ i } })-{ m }_{ i }J'_{ l }({ { k }_{ 0 }{ a }_{ i } })J_{ l }({ { k }_{ i }{ a }_{ i } }) },
		\label{eq12}
		\end{equation}
where the primes denote derivatives of the functions with respect to their arguments, and the superscripts \emph{s} and \emph{p} denote transverse electric and magnetic wave, respectively. Also, the elements of vector $[A]$ are given by	
		\begin{equation}
		{ a }_{ i }^{ l }={ e }^{ j{ k }_{ 0 }{ \rho  }'_{ i }\cos { ({ \phi ' }_{ g }-{ \phi  }_{ 0 }) }  }{ j }^{ l }{ e }^{ -jl{ \phi  }_{ 0 } },
		\label{eq13}
		\end{equation}	
where and the vector $[C]$ represents the unknown expansion coefficients of  scattered fields from the \emph{M }cylinders, which can be solved numerically. Then the scattering coefficient can be obtained by

		\begin{equation}
		F(\alpha )=-\sum _{ g=1 }^{ M }{ { e }^{ jk{ \rho  }'_{ g }\cos { ({ \phi ' }_{ g }-\alpha ) }  }\sum _{ n=-\infty  }^{ \infty  }{ { c }_{ gn }{ j }^{ n }{ e }^{ jn\alpha  } }  }.
		\label{eq14}
		\end{equation}
		\begin{equation}
		\sigma _{ ext }^{ s,p }=\frac { 2\lambda  }{ \pi  } Re(F(\alpha =0))
		\label{eq15}
		\end{equation}	
According to the optical theorem\cite{bohren2008absorption}, the extinction cross section for the whole coupled cylinders system can be retrieved by Eq.\ref{eq15}. Consequently, we can analyze the coupling effect between resonance modes, as well as the scattering contribution of each resonance mode by the coupling matrix $[S]$ shown in Eqs. \ref{eq8} and \ref{eq9}.

\section{Results and discussion}\label{sec3}

Strong toroidal dipole resonances have been found in compact clusters consisting of all-dielectric nano-cylinders\cite{PhysRevX.5.011036,TTKS16}, which can notably enhance EM fields in the free-space. To actively adjust toroidal dipolar resonances and further apply such resonant mode in light-matter interactions, this work is dedicated to proposing a silicon metamolecule with broad tunable frequency ranges and strong energy confinement inside the dielectric material instead of the free-space. Meanwhile, the coupling mechanism of collective resonance modes is elaborately investigated for the propose of active and more efficient design of the advanced nanoscaled optical devices.  
\begin{figure}[htbp]
	\centering
	\includegraphics[keepaspectratio,width=0.7\textwidth,height=0.7\textheight]{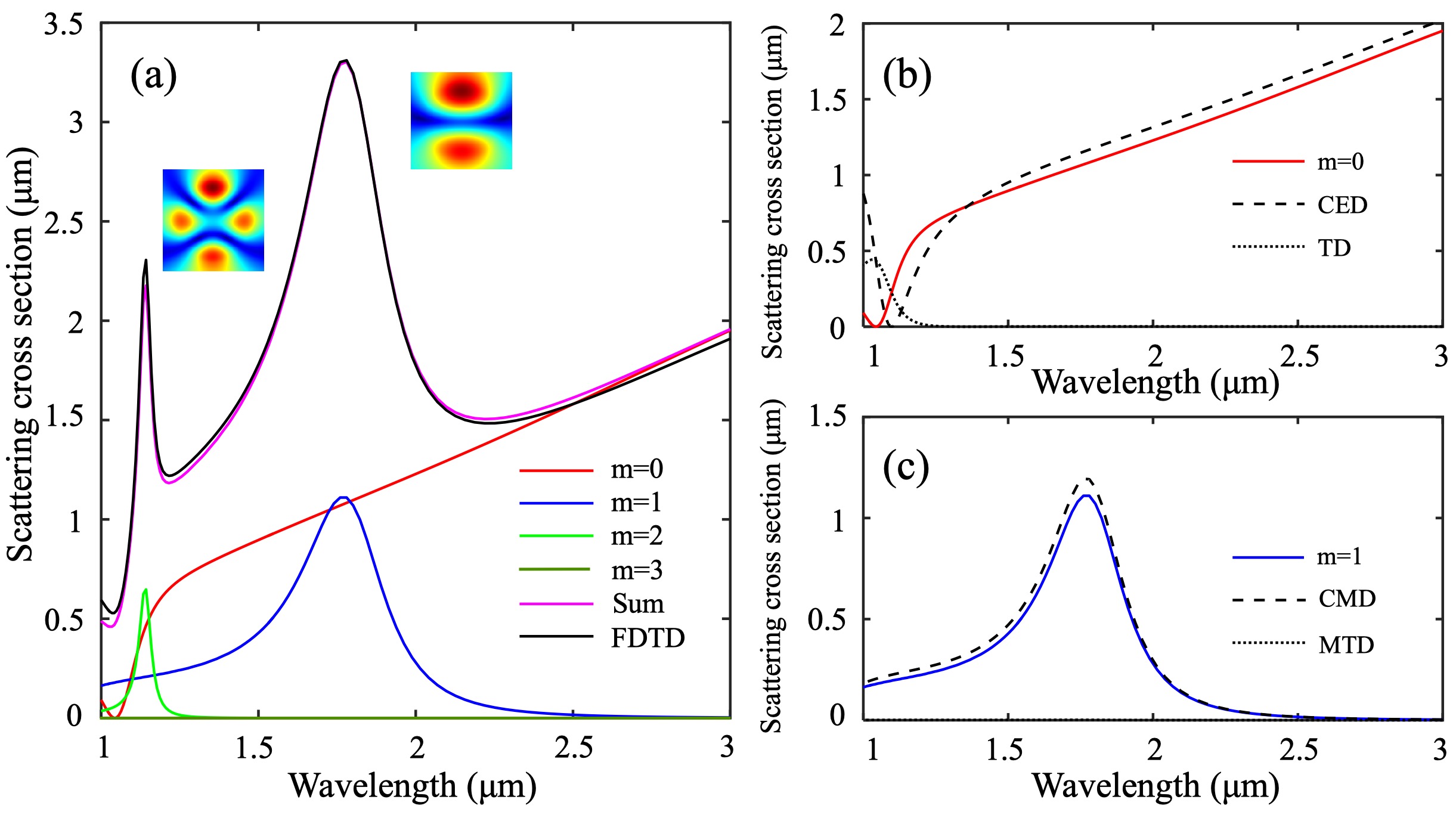}
	\caption{Decomposition of the scattering cross section of a cylinder with radius=190 nm in the 2D multipole expansion with s-polarized plane wave excitation. Plane (a) shows the contribution of each resonance mode. Inset of (a) shows electric distributions of the corresponding resonance peaks. The expanded toroidal dipole modes and Cartesian dipole terms are shown in (b) and (c), respectively.}\label{fig3a}
\end{figure}

As a reference case of our following investigation, we apply Cartesian multipole decomposition method to show the scattering spectra of an isolated cylinder with radius=190 nm, which is excited by s-polarized plane waves. From figure \ref{fig3a} (a), we can find that the overall scattering cross section from multipole decomposition method agrees well with FDTD simulation results, which verifies the multipole decomposition method. Also, we observe that two resonances caused by m=1 (MD) and 2 (MQ), when their electric fields are shown in the inset. Figs. \ref{fig3a} (b) and (c) respectively present that the TD mode has noticeable contribution when wavelength is shorter than 1.2 {$\upmu$}m, and the influence of MTD is negligible. Thus, we conclude that the toroidal response in an isolated cylinder is weak until a multi-cylinders metamolecule is proposed.

Our investigations about the tunability of TD modes in metamolecules are conducted as follows. Effects of cylinder separation $D$ on resonance frequencies, energy distributions, and Q factor are exhibited in Sec.\ref{sec3_1}. Then we discuss the effects of central cylinder in Sec.\ref{sec3_2} and give overall illustrations of the resonance frequencies and energy density with different $R_c$. Finally in Sec.\ref{sec3_3}, analyses on coupling mechanism of TD modes are estimated by the multiple scattering theory.   

\subsection{Effects of cylinder separation $D$}\label{sec3_1}
\begin{figure}[htbp]
	\centering
	\includegraphics[keepaspectratio,width=0.6\textwidth,height=0.6\textheight]{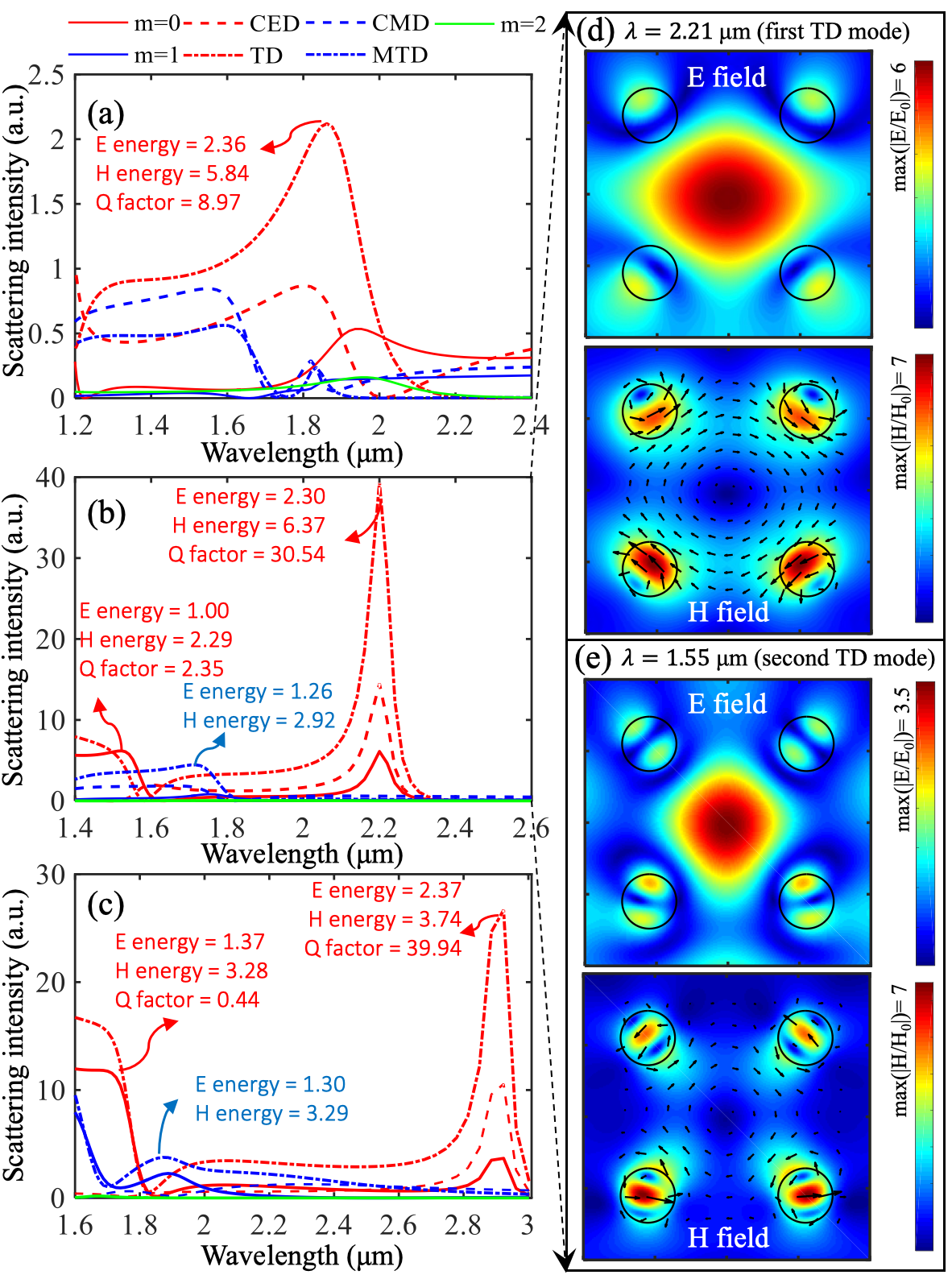}
	\caption{Multipole decomposition of the scattering intensity of quadrumer metamolecules with $R_c$=0, $R_s$=190 nm, $D$=600 (a), 1100 (b), 1600 (c) nm, and their corresponding electric and magnetic energy distributions. Q factor of each TD mode is also exhibited. EM fields of TD modes in the metamolecule with D=1100 are shown in (d) and (e), where outlines of cylinders are indicated by black circles and magnetic field vectors are denoted by black arrows.}\label{fig3}
\end{figure}

According to previous study we can find that TD modes can be affected by cylinder separation $D$. Here we use Cartesian multipole decomposition method to investigate multiple modes in the metamolecule (shown in Fig. \ref{fig1}) with $D$= 600, 1100, and 1600 nm, when $R_c$=0 and $R_s$=190 nm. From Fig.\ref{fig1}, coordinate positions of the surrounding cylinders are defined as $(x,y)=(\pm D/2, \pm D/2)$. Not only the scattering spectra are shown in Fig. \ref{fig3}, but also EM energy inside cylinders, which are calculated by $\int{{ \epsilon  }_{ 0 }{ \epsilon  }_{ r }{ \left| \textbf{ E }  \right|  }^{ 2 }ds}$ and $\int{{ \mu  }_{ 0 }{ \mu  }_{ r }{ \left| \textbf{ H }  \right|  }^{ 2 }ds}$. In addition, Q factor of each TD mode is also exhibited in Fig. \ref{fig3}. 
From Figs. \ref{fig3} (a), (b), and (c), we can observe that m=0 and 1 modes play the leading roles in the frequency range, where one or two TD modes can be distinctly excited. With the increment of cylinder separation $D$, we can find that one TD (the first TD) mode experiences a red shift from 1.85 to 1.92 $\upmu$m, and another TD (the second TD) mode becomes more independent and explicit, which also has a red shift from 1.28 to 1.75 $\upmu$m. Besides, we observe that Q factor in the first TD increases with $D$, when Q factor of the second TD decreases with $D$. Except from TD modes, a magnetic dipolar resonance can also be noticed in Figs. \ref{fig3} (b), and (c). After comparing the EM energy distributions of these resonances, we can notice that the first TD can localize more energy than the second mode, and the EM energy in the first TD mode reach its maximum when $D$=1100 nm. Thus, we calculate and discuss the EM fields in both TD modes when $D$=1100 nm, which are respectively shown in Figs. \ref{fig3} (d) and (e). Note that in each TD mode, a magnetic loop and a localized electric field in the center of metamolecule can be found, which confirm the appearance of toroidal dipolar resonances and result in the electric energy is smaller then the magnetic energy inside cylinders. Additionally, EM fields of the second TD are less symmetrical since this TD mode is overlapped with the adjacent magnetic dipolar resonance.
   
Adjusting the cylinder separation offers a method to tune the TD modes in the proposed metamolecules, but this method is too general and brings an overall influence to all resonance modes including the collective magnetic dipolar response. Besides, the energy confinement ability drops when $D$ is too large. Thus, as a primary step in the design, a carefully selected $D$ is required to provide relative strong and distinct toroidal dipole resonances.    

\subsection{Effects of the central cylinder}\label{sec3_2}
\begin{figure}[htbp]
	\centering
		\includegraphics[keepaspectratio,width=0.6\textwidth,height=0.6\textheight]{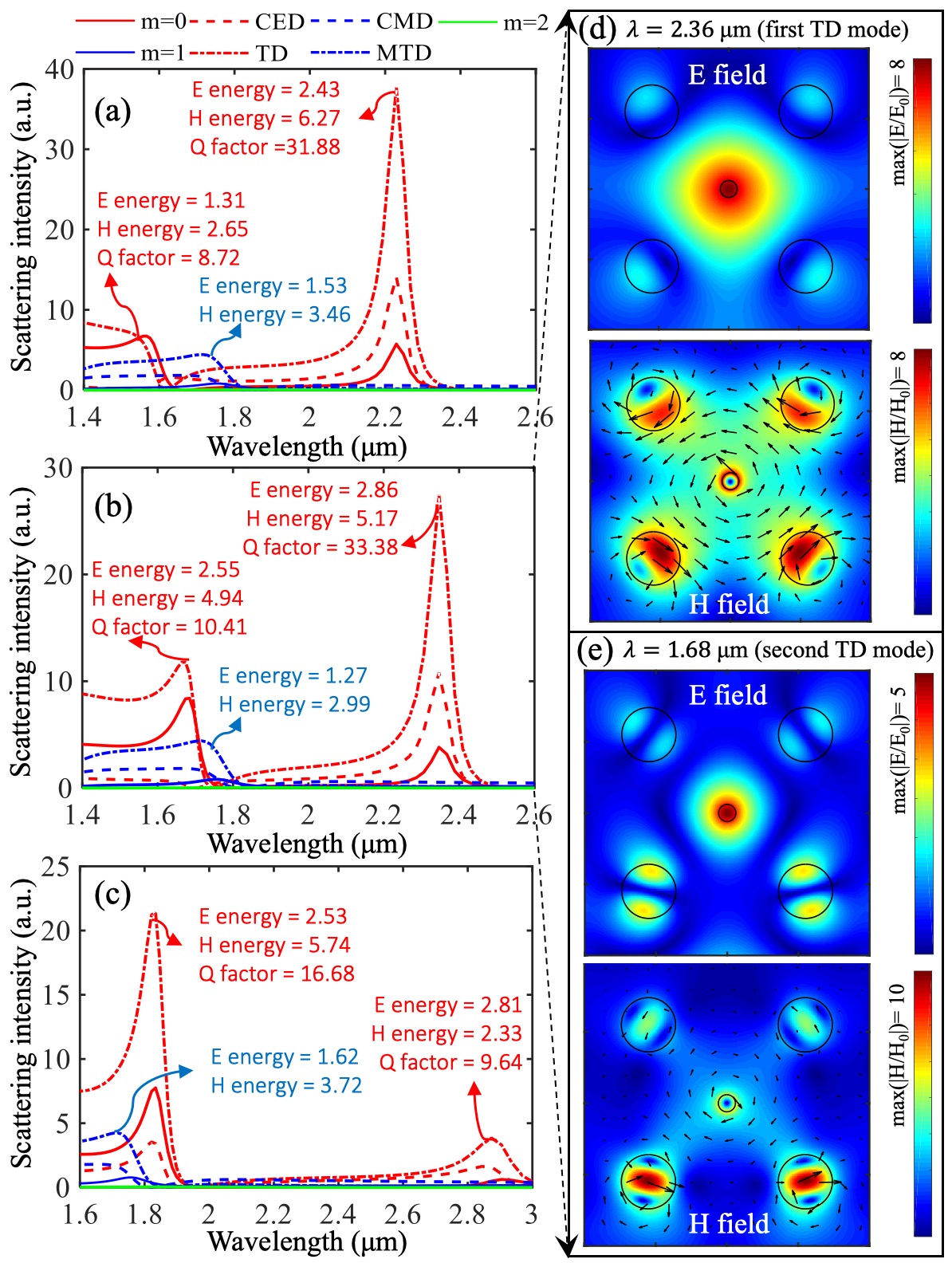}
	\caption{Multipole decomposition of the scattering intensity of pentamer metamolecules with  $R_s$=190 nm, $D$=1100 nm, $R_c$=30 (a), 60 (b), 120 (c) nm, and their corresponding electric and magnetic energy distributions. Q factor of each TD mode is also exhibited. EM fields of TD modes in the metamolecule with $R_c$=60 nm are shown in (d) and (e), where outlines of cylinders are indicated by black circles and magnetic field vectors are denoted by black arrows.}\label{fig5}
\end{figure}

Here we continue to investigate the tuning of TD modes by changing the size of central cylinder. Pentamer metamolecules with $R_c\neq 0$ are investigated following the route shown in Sec.\ref{sec3_1}. The only variable in pentamer metamolecules is the radius of central cylinder \emph{R\textsubscript{c}}, when \emph{R\textsubscript{s}} =190 nm and D=1100 nm. The scattering spectral and EM energy distributions are shown in Figure \ref{fig5} when $R_c$=30, 60, and 120 nm. In addition, Q factor of each TD mode is also exhibited in Fig. \ref{fig5}.

As shown in Fig.\ref{fig5}, with the increasing $R_c$, both the first and second TD resonance peaks experience red shifts, when the collective magnetic dipolar response remains in a fixed wavelength (1.78 $\upmu$m). However, the scattering power and Q factor of both TD modes show different performances with the increasing $R_c$: the first TD resonance becomes sharper when $R_c$<= 60 nm and turns weak when $R_c$=120 nm; the scattering intensity of second TD resonance grows stronger and sharper all the while. Different from the results in Fig. \ref{fig3}, the second TD resonance can exceed the first TD mode when the metamolecule is tuned by $R_c$. The metamolecule with $R_c$=60 nm has highest electric energy and Q factor when other factors are controlled, thus we further plot EM fields of TD modes in this metamolecule, which are shown in Fig. \ref{fig5} (d) and (e). Compared with Fig. \ref{fig3}, the EM fields of TD modes in the metamolecule with $R_c$=60 nm are more confined and have larger maximum. The red shift of the second TD mode makes it more overlapped with the collective magnetic dipolar mode, which results in its EM fields less symmetrical than EM fields in Fig. \ref{fig3} (e). Furthermore, magnetic vortexes around the central cylinder and surrounding cylinders have the same direction in the first TD, while they have opposite directions in the second TD, which indicate an anti-bounding state. 
\begin{figure}[htbp]
	\centering
	\includegraphics[keepaspectratio,width=\textwidth,height=\textheight]{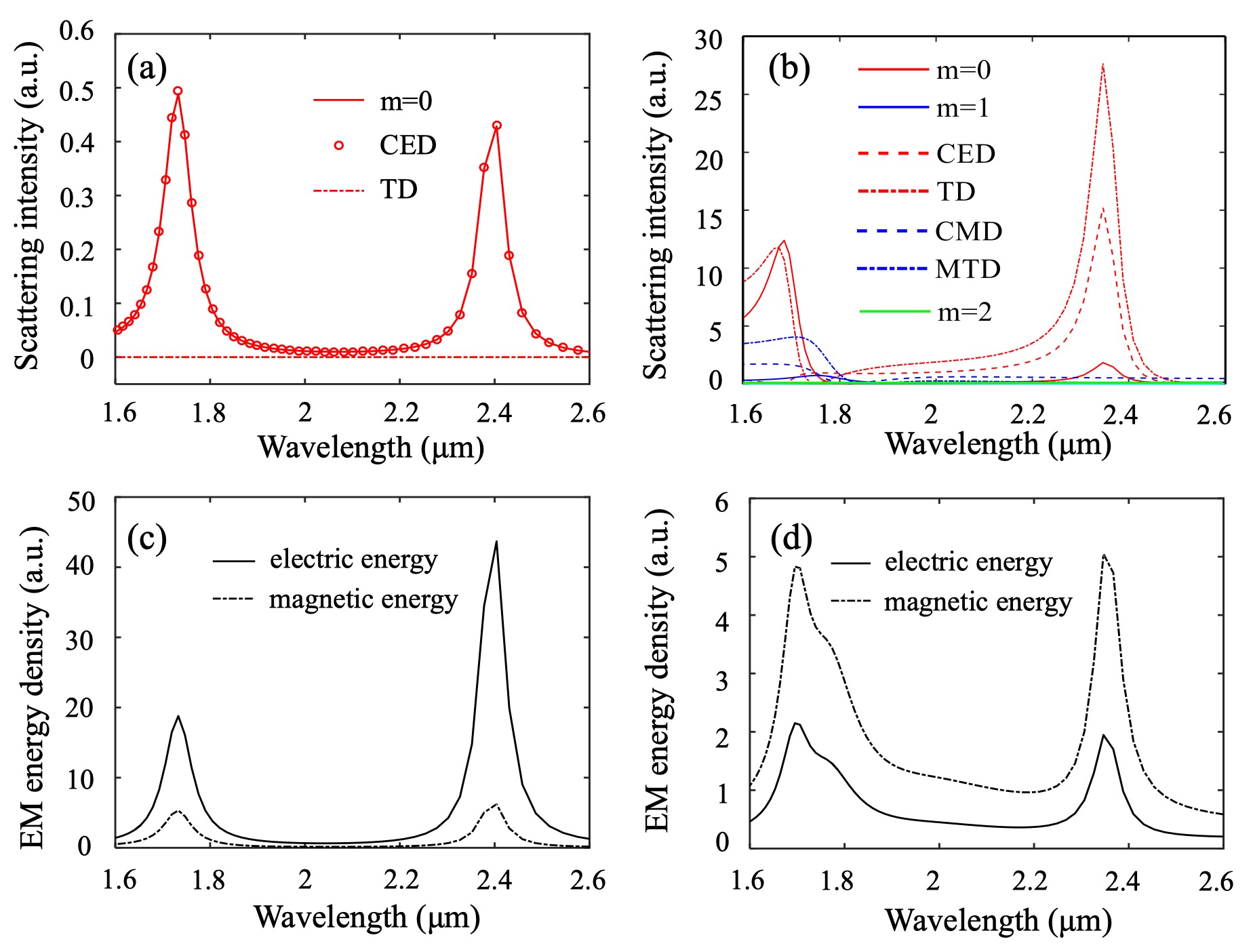}
	\caption{Multipole decomposition of the scattering intensity of the central cylinder (a) and surrounding cylinders (b) in pentamer metamolecules. Their EM energy density distributions are shown in (c) and(d), respectively. }\label{fig6}
\end{figure}

We further respectively investigate the resonance modes in the central cylinder ($R_c$=60 nm) and the surrounding cylinders, along with the normalized EM energy density. As seen in Figs. \ref{fig6} (a) and (c), the only supported resonance mode in the central cylinder is the electric dipole, resulting a dominate electric energy over the magnetic one. Scattering spectra of the surrounding cylinders and overall pentamer are similar, and the energy distributions in Figs. \ref{fig6} (c) and (d) show that the collective magnetic mode only brings energy enhancement in surrounding cylinders, which suggests this collective magnetic mode is only induced by the surrounding cylinders. It is also worthwhile to mention that the electric density in the central cylinder can reach 20 times than in the surrounding cylinders when the first TD mode is excited, which confirms the central cylinder can provide a strong electric field enhancement in the dielectric material instead of free space.
\begin{figure}[htbp]
	\centering
	\includegraphics[keepaspectratio,width=\textwidth,height=0.75\textheight]{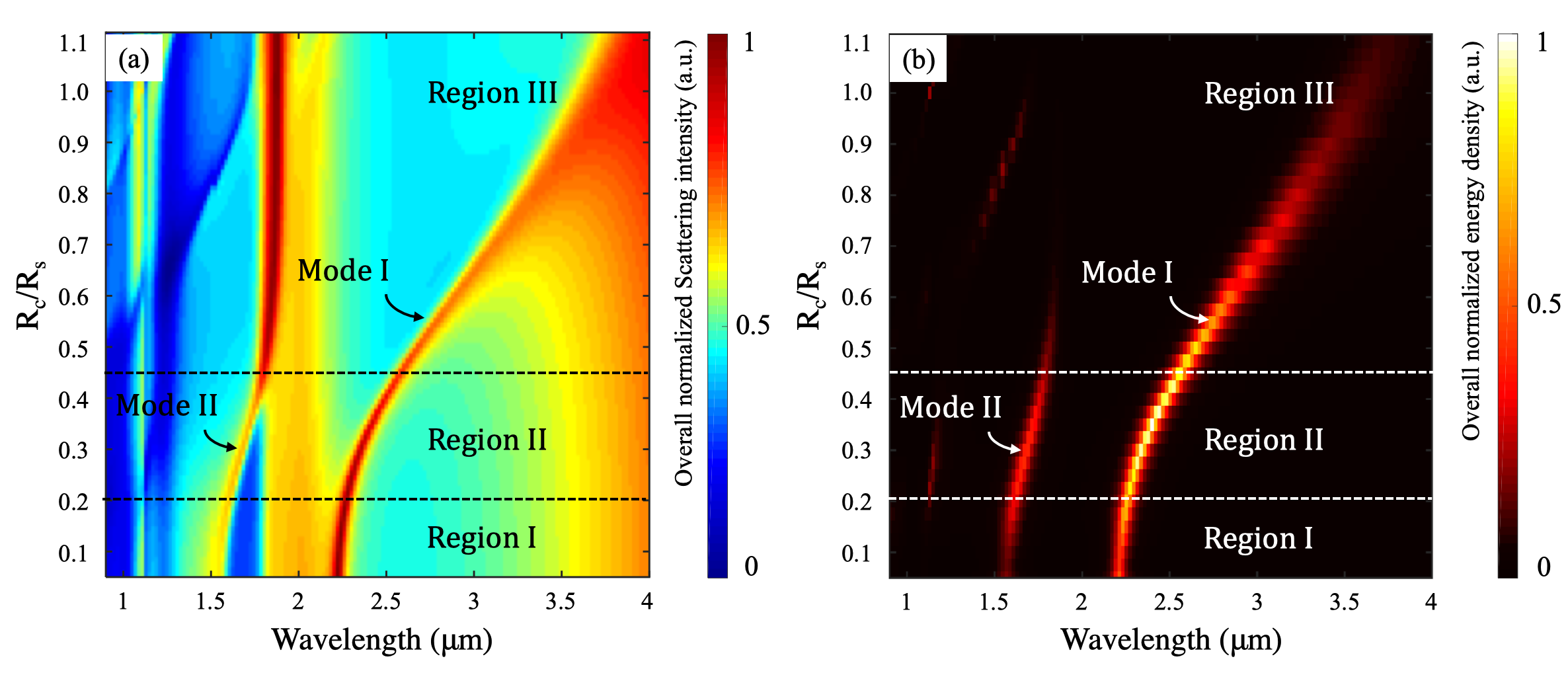}
	\caption{Spectral scattering intensity of metamolecules (a) and electric energy density of the central cylinder (b) with the changing of $R_c/R_s$.}\label{fig7}
\end{figure}

To intuitively display the effect of $R_c$, scattering spectra of metamolecules and electric energy density distributions of the central cylinder with the changing of $R_c/R_s$ are shown in figs \ref{fig7} (a) and (b), respectively. Compared with Fig. \ref{fig5}, the two distinct scattering bands in both (a) and (b) are corresponding to the first and second TD mode, which are marked by Mode 1 and 2, respectively. The red shifts of both TD are clearly shown in figure (a), according to which we could roughly divide the y axis into three regions: (i) both TD modes can hardly be affected by the central cylinder in Region I; (ii) obvious red shifts of both TD modes can be noticed in Region II; (iii) red shift only happens to the first TD mode, when the second TD mode remains in a fixed wavelength in Region III. Even though Fig. \ref{fig7} (a) shows a clearly evolution of both TD modes, it can hardly provide the energy confinement properties of metamolecules. Fig. \ref{fig7} (b) shows the electric energy density of the central cylinder, where the maximum electric intensity appears according to previous discussion. It is obviously that the first TD mode is always accompanied with a larger energy density than the second TD mode. Besides, the electric energy density of the first TD mode is mostly enhanced in Region II, and falls gradually when ($R_c/R_s$) moves away from this Region. Similarly, the strongest energy density enhancement induced by the second TD mode is also occurs in Region II. Thus, we can easily adjust the resonance frequencies and their corresponding field enhancement by tuning the ($Rc/Rs$) ratio. The tunable region of the first TD mode is ranging from 2.21 $\upmu$m to over 3.55 $\upmu$m, while the second TD mode has a smaller tunable frequency band ranging from 1.53 $\upmu$m to 1.84 $\upmu$m.

Compared with changing the cylinder separation $D$, adjusting the size of central cylinder can not only provide the tuning of both TD modes without influences on the collective magnetic dipolar resonance, but also modulate the relative strength of both TD modes. For instance, we can make the first TD mode stronger in Region I and II, or realize a larger second TD resonance in Region III. From the analyses above, we can preliminarily understand the tunability of the central cylinder is due to an additional electric dipolar response. However, the underlying mechanism of modes coupling in these toroidal resonances still requires further discussion.

\subsection{Analyses of coupling mechanism in TD resonances}\label{sec3_3}
\begin{figure}[htbp] 
	\centering
	\includegraphics[keepaspectratio,width=\textwidth,height=0.5\textheight]{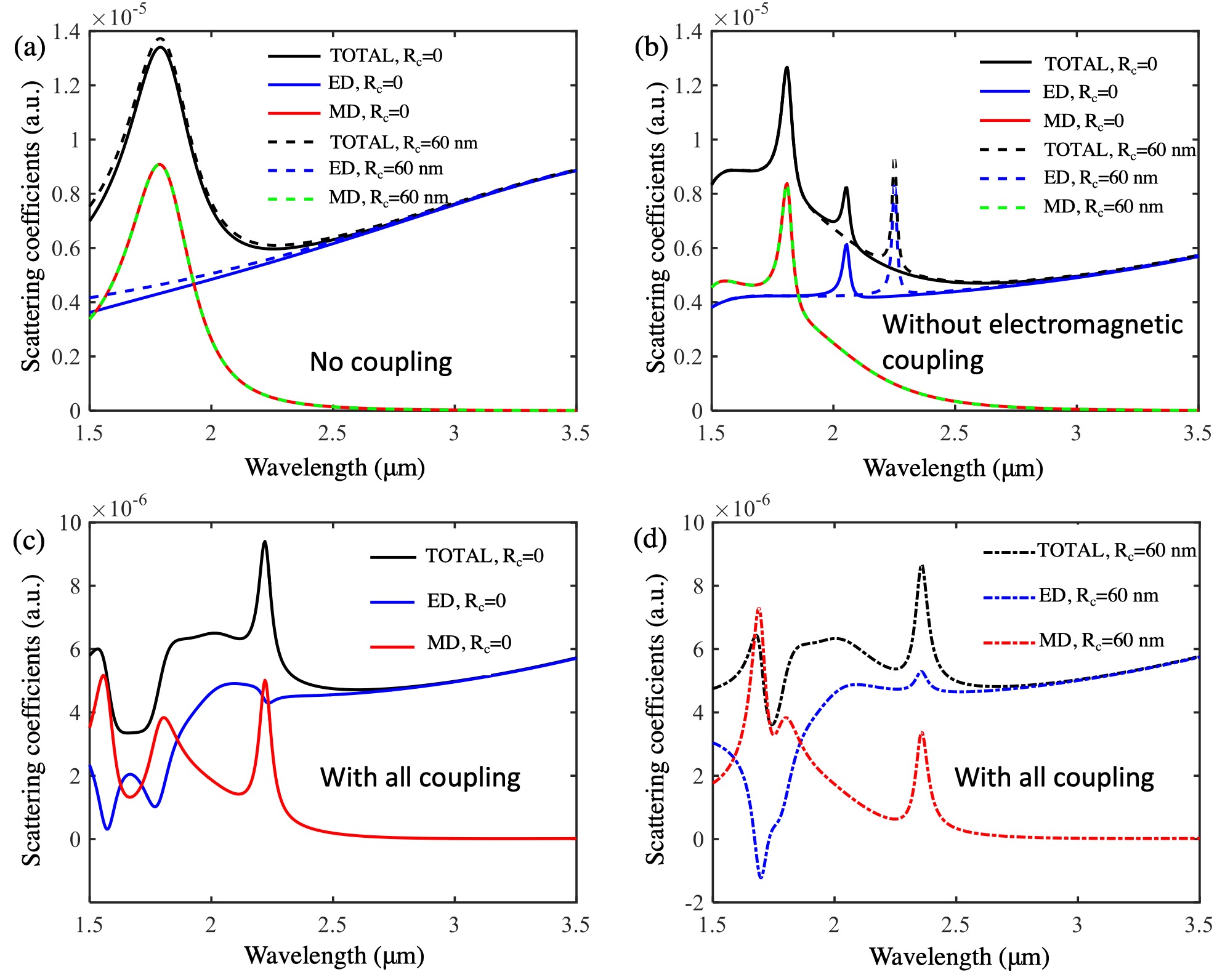}
	\caption{Coupling effects on the scattering coefficients spectra of the metamolecules with $R_c$=0 and 60 nm, when $R_s$=190 nm, $D$=1100 nm: (a) considers no coupling effect; (b)  neglects the cross electromagnetic coupling; all electromagnetic couplings are considered when $R_c$=0 (c) and 60 nm (d).}\label{fig8}
\end{figure}
\begin{figure}[htbp]
	\centering
	\includegraphics[keepaspectratio,width=0.6\textwidth,height=0.6\textheight]{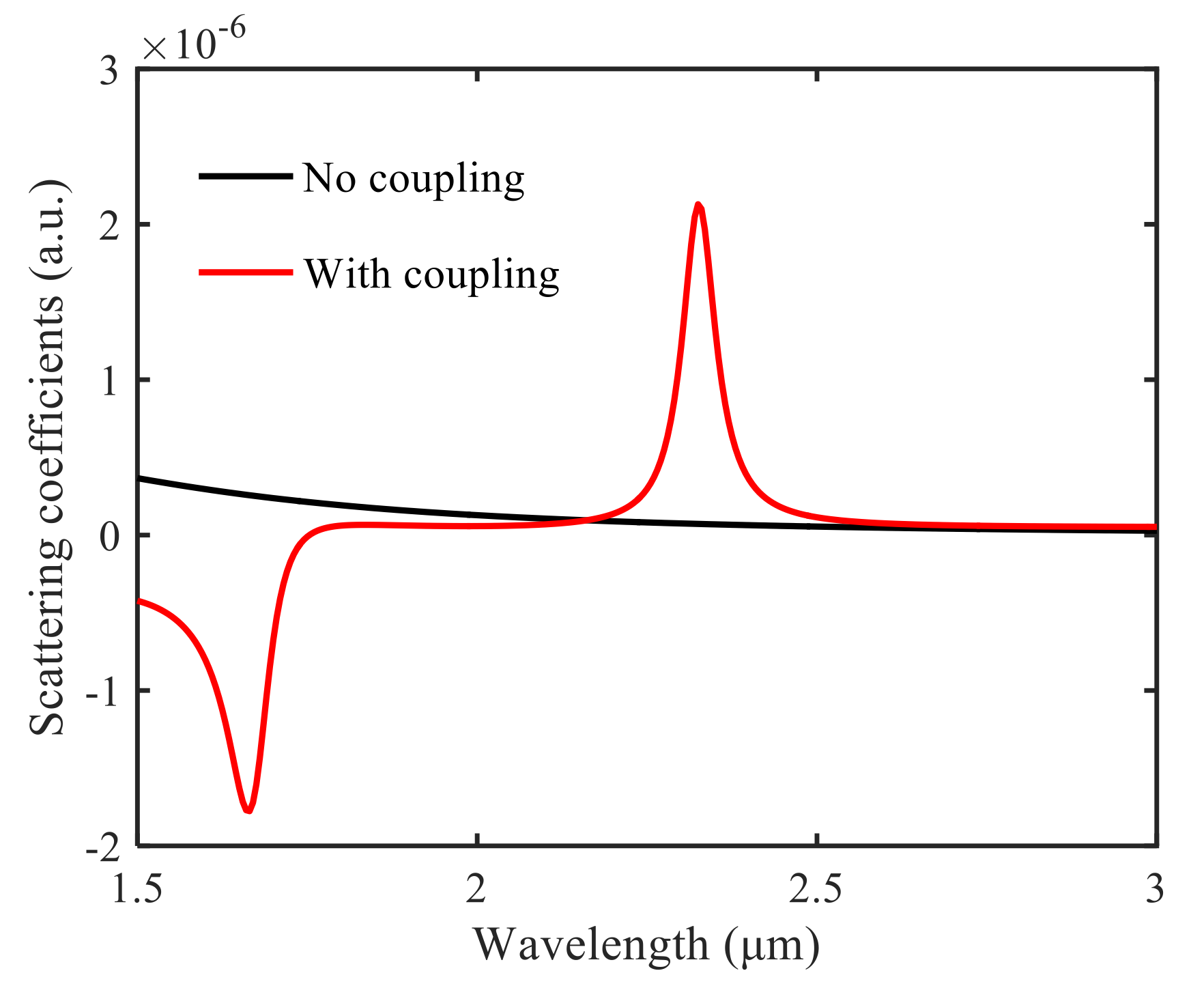}
	\caption{Coupling effects on the scattering coefficients spectra of the central cylinder when $R_c$=60 nm, $R_s$=190 nm, and $D$=1100 nm.}\label{fig10}
\end{figure}
To further reveal the underneath mechanism of modes coupling inside the pentamer, we apply the multiple scattering method to provide a theoretical analysis in the view of conventional spherical coordinate, which is adaptable since the toroidal dipole can always be regarded as the head-to-tail aligned magnetic dipoles. In this part, three coupling situations of metamolecules with $R_c$=0 and 60 nm are considered, which are: (i) consider no coupling effect by setting $S_{ig}^{ln}$=0 when $i\neq{g}$ or $l\neq{n}$; (ii) consider only the contributions from both EDs and magnetic multipoles but neglect the cross electromagnetic coupling by setting  $S_{ig}^{ln}$=0 when $l\neq{n}$ and $l\times{n}=0$; (iii) consider all modes coupling. As illustrated in Fig \ref{fig7} (a), $R_c$ only affects the electric dipolar response of metamolecules when no coupling is considered, and a magnetic resonance peak emerges in 1.78 $\upmu$m, which is the instinct MD dipole mode in each surrounding cylinder with $R_s$=190 nm. From Fig. \ref{fig7} (b), we can find that the coupling between EDs or MDs can respectively induce a sharp peak, when the magnetic one is not affected by $R_c $ and remains in 1.78 $\upmu$m. After considering all modes coupling in Figs \ref{fig7} (c) and (d), we can find that the induced collective EDs mode, which is shown in Fig. \ref{fig7} (b), results in a peak of magnetic dipolar in 2.21 (2.37) $\upmu$m when $R_c$=0 (60) nm, which is known as the first TD mode. According to this fact, we can easily tune the first TD mode by the additional electric dipolar response offered by the central cylinder. However, no resonance peak corresponding to the second TD is found until we take cross electromagnetic coupling into account, which suggests that the second TD is resulted from the interference of electric and magnetic dipolar resonances. Thus, to tune the second TD mode requires both ED and MD responses. As we know that the magnetic response in the proposed metamolecule mainly comes from the collective magnetic mode ($\lambda=1.78 \upmu$m), which is independent of $R_c$, and the central cylinder can only offer an additional electric dipolar tunability. Due to their different formation mechanism, the second TD, not surprisingly, has a narrower tunable frequency band than the first TD mode. 

Moreover, we calculate the partial scattering coefficients of the central cylinder in the pentamer with $R_c$=60 nm. Compare Figs. \ref{fig8} (d) and \ref{fig10}, we can notice that the negative ED partial scattering coefficient in the metamolecule is caused by the negative ED scattering coefficient in central cylinder. From previous magnetic field vectors of the second TD, we can understand that the negative value is resulted by the opposite directions of the magnetic vortex current in the central and surrounding cylinders, which can be regarded as a partial destructive interference. Besides, we can observe that a interference between the second TD and the collective magnetic dipole mode creates a dip in scattering spectra, which on the contrary has a relatively high energy as shown in Fig.\ref{fig5} (d). The Fano resonance \cite{Hopkins2015Interplay,Chong2014Observation,Lassiter2012Designing} resulted from the coupling of conventional electric and magnetic dipole can be seen as an analogy of this phenomenon, since a TD shares same far-field scattering pattern with an electric dipole.

\section{Conclusion}\label{sec4}
In summary, we investigate the scattering of normally incident s-polarized plane waves by the pentamer metamolecules through Cartesian multipole decomposition method where toroidal family has been considered. It is shown that two distinct TDs can be excited within homogeneous dielectric nano-cylinders, when cylinder separation $D$ is carefully selected. The central cylinder, which provides only an electric dipolar response, can efficiently tune TD resonances in their frequencies and energy confinement. The coupling matrix in multiple scattering theory is used to detailed investigate the coupling mechanism of toroidal dipolar resonances. In this way, we find that the first TD is mainly caused by the coupling of electric dipole modes from all cylinders, which then induces head-to-tail aligned magnetic dipole rings. However the second TD mode requires the cross coupling between electric and magnetic dipolar resonances, which means it need additional electric and magnetic dipolar responses, simultaneously. Thus, the first TD has broader tunable frequency range than the second TD, since the central cylinder can only provide a changeable electric dipolar response. Here we believe this research on a low-loss metamolecule with two highly tunable toroidal dipolar resonances may offer new insights into the understanding of toroidal modes in multiple nano-cylinders system, and hopefully can offer a new platform for active designing in sensing, nanoantennas, lasers, and photovoltaic devices.

\section*{Acknowledgement}
This work is supported by the National Natural Science Foundation of China (Grant Nos.51636004, 51476097), Shanghai Key Fundamental Research Grant (No.18JC1413300, No.16JC1403200) and the Foundation for Innovative Research Groups of the National Science Foundation of China (Grant No. 51521004).              
\bibliography{sample}






\end{document}